\documentclass[twocolumn,secnumarabic, aps, prl]{revtex4}

\usepackage{hyperref}

\usepackage{amssymb,amsmath,amsfonts}
\usepackage{epsfig}
\usepackage{array}
   \usepackage{graphicx}

\usepackage{amsmath}
\usepackage{multirow}
\usepackage{array}

\newcommand{\nn}{\nonumber}

\begin{document} 

\title{  Exciton Condensation in a Holographic  Double Monolayer  Semimetal }

\author{Gianluca Grignani,  Andrea Marini }
\affiliation{ Dipartimento di Fisica e Geologea, Universit\`a di Perugia,
INFN Sezione di Perugia, 
Via A. Pascoli, 06123 Perugia, Italia}
\author{ Namshik Kim, Gordon W. Semenoff }
\affiliation{Department of Physics and Astronomy, University of British Columbia, 
                     6224 Agricultural Road, Vancouver, BC, Canada V6T 1Z1}


\begin{abstract}
The formation of intra-layer and inter-layer exciton condensates in a model of a double monolayer Weyl semi-metal 
is studied in the strong coupling limit using AdS/CFT duality.  We find a rich phase diagram which includes phase transitions between inter-layer and intra-layer
condensates as the charge densities and the separation of the layers are varied.  The tendency to inter-layer
condensation is strongest when the charge densities are balanced so that the weak coupling 
electron and hole Fermi surfaces would be  nested.   For systems with multiple species of massless fermions, 
we find a novel phase transition where the charge balance for nesting  occurs spontaneously. 
\end{abstract}

\maketitle

The possibility that an inter-layer exciton condensate can form in a double monolayer of two-dimensional electron gases 
has been of interest for a long time \cite{lozovik}.  A double monolayer
contains two layers, each containing an electron gas, 
separated by an insulator so that electrons cannot be transferred between the layers.
Electrons and holes in
the two layers can still interact via the Coulomb interaction.  The exciton  which would condense is a bound
state of an electron in one layer and a hole in the other layer.  This idea has
recently seen a revival with some theoretical computations for emergent relativistic 
systems such as graphene or some topological
insulators which suggested that a
condensate could form  at relatively high temperatures, even at room temperature \cite{graphene}.  
A room temperature superfluid would have applications in electronic devices where proposals include ultra-fast switches 
and dispersionless field-effect transistors \cite{device}.    

An exciton condensate might be more readily achievable in a double monolayer with relativistic  electrons
due to particle-hole symmetry and the possibility of engineering nested Fermi surfaces of electrons in one layer and the holes in the other layer.  
This nesting would 
enhance the effects of the attractive Coulomb interaction between an electron and a hole.  Even at very weak coupling, it can be shown to
produce an instability to exciton condensation  \cite{babak}. 
However, in spite of this optimism, an inter-layer condensate has yet to be observed in a relativistic material, even in 
experiments using clean graphene sheets with separations down to the nanometer scale \cite{nano}.  The  difficulty with theoretical computations,
where the Coulomb interaction is strong, is the necessity of ad-hoc inclusion of screening, 
to which the properties of the strongly coupled system have been argued to be sensitive \cite{screen}.

\begin{center}
\begin{figure}
	 \includegraphics[scale=1]{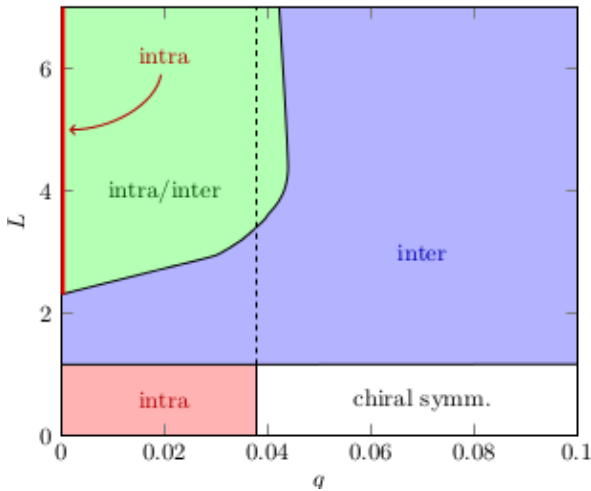}
	 	\caption{(color online) 
	Phase diagram of the charge balanced double monolayer (exactly nested Fermi surfaces).  The vertical axis is layer separation $L$ in units of
	the inverse ultraviolet cutoff, $R$.
	The horizontal axis is the charge density $q$ in units of $R^{-2}$.   
	The green region has both inter- and intra-layer condensates.  The blue region has only an inter-layer condensate.
	The red region has only an intra-layer condensate.  
	The white region has no condensates. \label{phase} }
	\end{figure}
	\end{center}

In this paper we will study a model of a double monolayer of relativistic two-dimensional electron gases.   
This model has a known AdS/CFT dual which is easy to study and it can be solved exactly
in the strong coupling limit.   
We shall learn that, in this model. the only condensates which form 
are excitons, bound states of electrons with holes in the same layer (intra-layer)
or bound states of electrons in one layer with holes in the other layer (inter-layer).
Moreover, even though at very strong coupling, the idea of a Fermi surface loses its meaning, we find that 
the tendency to form an inter-layer condensate is indeed greatly enhanced by the charge balance which, at weak coupling,
would give nested particle and hole Fermi surfaces.   We shall see that, in the strong coupling limit, 
and when the charges are balanced, an inter-layer condensate can form for any separation of the 
layers.  As well as the inter-layer condensate, such a strong interaction will also form an intra-layer condensate.   We find
that a mixture of the two condensates is favoured for small charge densities and larger layer separations.  
For sufficiently large charge densities, on the other hand, the only condensate is the inter-layer condensate.  
These results for charge balanced layers 
are summarized in figure~\ref{phase}.
When the charges are not balanced, so that at weak coupling the Fermi surfaces would not be nested, 
no inter-layer condensate forms, regardless of the layer separation.   This dramatic difference is similar to and even sharper than
what is seen at weak coupling \cite{babak} where condensation occurs in only a narrow window of densities near nesting.

However, even in the non-nested case, we can find a novel symmetry breaking 
mechanism where an inter-layer condensate can form.   
If each electron gas contains more than one species of relativistic electrons (for example, graphene has
four species of massless Dirac electrons and some topological insulators  have two species), 
the electric charge can
redistribute itself amongst the species to spontaneously nest one or more pairs of Fermi surfaces, with the unbalanced
charge taken up by the other electron species.   Then the energy is lowered by formation of a condensate of the nested
electrons, the others remaining un-condensed.  To our knowledge, this possibility has not been studied before.
The result is a new kind of symmetry breaking where Fermi surfaces nest spontaneously and break some of the internal symmetry
of the electron gas in each layer.  We demonstrate that, for some examples of the charge density, this type of condensate indeed
exists as the lowest energy solution.

The model which we shall consider is a defect quantum field theory consisting of a pair of parallel, 
infinite, planar 2+1-dimensional defects in 3+1-dimensional Minkowski space  and separated by a distance $L$.    
The defects are each inhabited by $N_F$  
species of relativistic massless Dirac fermions.   The fermions interact by exchanging massless quanta
of maximally supersymmetric Yang-Mills theory which inhabits the surrounding 3+1-dimensional bulk. 
In the absence of the
defects, the latter would be a conformal field theory. The interactions which it mediates have a $1/r$ fall-off, similar to the Coulomb 
interaction and, in the large $N$ planar
limit which we will consider,  like the Coulomb force, 
the electron-hole interaction is attractive in all channels.  
The field theory action is 
\begin{align}
S=&\int d^4x\frac{1}{g_{\rm YM}^2}{\rm Tr}\left[-\frac{1}{2 }F_{\mu\nu}F^{\mu\nu}-\sum_{b=1}^6D_\mu\Phi^b D^\mu\Phi^b+\ldots\right]   \nonumber \\
&+\int d^3x\sum^2_{a=1} \sum_{i=1}^{N_F}\bar\psi_{ai}\left[  i\gamma^\mu\partial_\mu + \gamma^\mu A_\mu  +  \Phi^6\right]\psi_{ai} 
\label{model}
\end{align}
The first term is the action of ${\cal N}=4$ supersymmetric Yang-Mills theory 
where   $A_\mu$ is the Yang-Mills gauge field and $\Phi^6$ is one of the scalar fields and the second term is the action of the defect
fermions. 
In the second term, the subscript $a$  labels the defects and $i$  the fermion species.  
The action includes all of the marginal operators which are compatible with the symmetries.  
It has a global U(1) symmetry which we associate with electric charge.
 
 The defect field theory (\ref{model}) is already  interesting  with one layer.   
It is thought to have a conformally symmetric weak coupling phase for $0\leq\lambda\leq\lambda_c$.
 When $\lambda>\lambda_c$,  
chiral symmetry is broken by an intra-layer exciton condensate \cite{Kutasov:2011fr}.  
Near the critical point,  the order parameter is thought to
scale as  $\left<\bar\psi_{1i}\psi_{1i}\right>\sim \Lambda^2\exp\left(-b/\sqrt{\lambda-\lambda_c}\right)$ where $\Lambda$ is an ultraviolet (UV)
cutoff.  In the strong coupling phase, the condensate and therefore the charge gap are finite only when the coupling is tuned to be close to its
critical value.  The holographic construction examines this theory in the strong coupling limit, where $\lambda\gg \lambda_c$.  In that limit, it is cutoff dependent
and it can only be defined by introducing a systematic UV cutoff.  We will find a string-inspired way to do this, tantamount to defining the model \eqref{model}
 as a limit of the IIB string theory which is finite and resolves the singularities.  It will allow us to study the strong coupling limit using the string theory dual
 of this system.  
 
  When there are two monolayers, the field theory (\ref{model}) can also have an inter-layer
exciton condensate with order parameter $\left<\bar\psi_{1i}\psi_{2i}\right>$.   The results of reference \cite{babak} suggest
that, with balanced charge densities and nested Fermi surfaces, the inter-layer condensate occurs even for very weak coupling.  
Not much is known as to how it would behave at strong coupling.  
It is the strong coupling limit of this model which we will now solve using its string theory dual.

The string theory dual of the defect  field theory  is the D3-probe-D7 brane system of IIB string theory\cite{rey}. 
 A monolayer is a single stack of $N_F$ D7 coincident branes.  A double monolayer
has two parallel stacks, one of $N_F$ D7 branes and another of $N_F$ anti-D7 branes separated by a distance $L$.
In both cases, the D7 brane stacks overlap  $N>>N_F$ coincident D3 branes.  
With the appropriate orientation, the lowest energy states
of  open strings which connect the D3 to the D7 branes are massless 
two-component relativistic fermions that propagate on  2+1-dimensions and are the defect fields
in (\ref{model}). In the large $N$ and strong coupling limits, the D3 branes are  replaced by 
the $AdS_5\times S^5$ background and solving the theory reduces to extremizing the classical Born-Infeld action
$S\sim N_FT_{\rm D7}\int d^8\sigma\sqrt{-\det(\gamma+2\pi\alpha' F)}$
for the D7 brane embedded with world-volume gauge field strength F and metric $\gamma_{ab}$ in $AdS_5\times S^5$.    
However, there is an
immediate  problem with this setup.  Any D7 brane geometry
which approaches the appropriate D7 brane boundary conditions at the boundary of $AdS^5$ is unstable.  This is a reflection of the fact that
the strong coupling limit of the quantum field theory on a single D7 brane is not conformally symmetric.  
We shall use a suggestion by Davis et.al. \cite{Davis:2008nv} who regulated the D7 brane by embedding it 
in the extremal black D3 brane geometry, with metric
\begin{align}
\frac{ ds^2}{R^2}& = \frac{ r^2 \left(-dt^2+dx^2+dy^2+dz^2\right)}{\sqrt{1+R^4r^4}}   \nonumber \\
&   +\sqrt{1+R^4r^4}\left(\frac{dr^2}{r^2}+ d\psi^2+\sin^2\psi \sum_{i=1}^5 (d\theta^i)^2\right) 
\label{metric}
\end{align}
where $\sum_{i=1}^5(\theta^i)^2=1$. 
 and $R^4=\lambda {\alpha'}^2$.  The asymptotic, large $r$ limit of this metric is 10-dimensional Minkowski space.   
It has a horizon at $r=0$.     In the near horizon limit, which produces the IIB string on $AdS_5\times
S^5$,  $rR\ll 1$, it approaches the Poincar\'e patch of $AdS_5\times S^5$.  
Since $R$ contains the string scale $\alpha'$, $1/R$ can be regarded
as a  (UV) cutoff. 

 The D7 and anti-D7 world-volumes are almost entirely determined by symmetry. They have
2+1-dimensional
Poincar\'e invariance and wrap $(t,x,y)$.  The model (\ref{model}) has an
$SO(5)$ R-symmetry.  The D7's must therefore wrap    $(\theta^1,\ldots,\theta^5)$ to form an $S^4$.
For the remaining world-volume coordinate, we use the radius $r$ in (\ref{metric}).  The dynamical variables
are then $\psi(r)$ and the positions $z_{1}(r)$ and $z_2(r)$ of D7 and anti-D7, which  by symmetry can only be
functions of $r$.  $\psi=\frac{\pi}{2}$ is a point of higher symmetry, corresponding to parity in the defect field theory with massless
fermions.  $\psi(r)=\frac{\pi}{2}+\frac{c}{r^2}+\ldots$ is required to approach $\frac{\pi}{2}$ at $r\to\infty$ and, if it becomes $r$-dependent at all ($c\neq 0$), parity   is broken by an intra-layer condensate.   Parity can be restored if pairs of branes have condensates of opposite signs. This would break
 flavour symmetry when $N_F$ is even,  $U(N_F)\to U(N_F/2)\times U(N_F/2)$. Whether this sort of flavour symmetry breaking or parity and time reversal
breaking takes place is an interesting dynamical question which will be studied elsewhere.
Finally, it will turn out that, either $z_{1,2}(r)$ are constants, or the D7 and anti-D7 meet and smoothly join together at a minimum radius, $r_0$. 
Asymptotically, $z_{1,2}(r)=\pm L/2 \mp R r_0^4/r^4+\ldots$.
 
 We have performed numerical computations to determine the lowest energy embeddings of the D7  (and anti-D7) branes as a function 
of the charge density ($q$) and the brane-anti-brane separation $L$.  In the following we outline the results of these computations. 
The formalism for studying the embeddings of the probe D branes is already well-known in the literature and we refer the reader there for details.
Examples for double monolayers can be found in references \cite{Davis:2011am}-\cite{new}. 
  
  When we suspend a single D7 brane in the black D3 brane metric (\ref{metric}), we find that 
the lowest energy solution  truncates before it reaches the  horizon. This is called a ``{\it  Minkowski embedding}''.  
The function $\psi(r)$ moves from  $\psi=\frac{\pi}{2}$  at $r\sim\infty$ to
$\psi=0$ or $\psi=\pi$ at the $r$ where the brane pinches off.  The $S^4$ which the world-volume wraps
shrinks to a point there and this collapsing cycle is what makes the 
truncation smooth.   
This brane geometry is interpreted as
a charge-gapped state.  The lowest energy charged excitation is a fundamental string which would be suspended between the D7
brane and the horizon.  In this case, that string has a minimum length and therefore a mass gap.    
 
We can introduce a charge density $q$ on the single monolayer.   
When the D7 brane carries a charge density, its world volume must necessarily
reach the  horizon.  This is called a ``{\it black hole embedding}''.   
Charge in the quantum field theory corresponds to  D7 brane world-volume 
electric field $E_r\sim q $. This hedgehog-like electric field points outward from the centre of the brane.
The radial lines
of flux of the electric field can only end if there are sources.   Such sources would be fundamental strings, suspended between the D7 brane
and the   horizon.   However, the strings have a larger tension than the D7 brane and they pull the the D7 brane to the horizon
resulting in a gapless state.  This is confirmed by numerical solutions of the embedding equation of a single brane and, indeed, we find
that the $S^4$ which is wrapped by the world volume shrinks to a point as it enters the   horizon.    
This state no longer has a charge gap.   
Even in the absence of a charge gap, we find that,
for small charge densities, there is still an intra-layer exciton condensate.  Our numerical studies show
that it  persists  up to 
a quantum phase transition  at a critical density $q_{\rm crit.}\approx  0.0377/R^2$.   At densities greater
than the critical one, $\psi=\frac{\pi}{2}$, is a constant.

Now, consider the double monolayer with D7 and anti-D7 branes.  
A D7-anti-D7 pair of branes would tend to annihilate.   We prevent this annihilation by requiring that they be separated by a distance $L$ as 
they approach the boundary at $r\to\infty$.  When their world volume enters the bulk, they can still come together and annihilate -- their
world volumes fusing together at a minimal radius $r_0$. This competes with the tendency of a monolayer brane to pinch off at some radius.  Indeed, when
the charge density is zero, we see both behaviours. When the stacks of branes are near enough, that is, $L<L_c\simeq 2.31 R$ is small enough, they join.  
This state has an
inter-layer condensate.  When they are farther apart, they  remain un-joined.  
Instead, they pinch off to form Minkowski embedding, corresponding to a state with intra-layer condensates. 

When we introduce balanced charges $q$ and $-q$ on the D7 and anti-D7, respectively, there are four  modes
of behaviour which are summarized in table \ref{typessolqnotzero}. Each of these behaviours occurs in 
the phase diagram in figure \ref{phase}.
 \begin{table}[!ht]
	\begin{center}
	{\renewcommand{\arraystretch}{1.1}
		\setlength{\extrarowheight}{2pt}
		\begin{tabular}{c|c|c|}
			\cline{2-3}
			& $z_2-z_1=L,~{\rm const.}$ & $z_2(r)-z_1(r)\to 0~{\rm at}~r_0$ \\
			\hline
			\multicolumn{1}{|c|}{\multirow{3}{*}{$c=0$}} & \textbf{Type 1} & \textbf{Type 2}\\
		 	\multicolumn{1}{|c|}{}	& un-joined, $\psi=\frac{\pi}{2}$ & joined, $\psi=\frac{\pi}{2}$\\
		  \multicolumn{1}{|c|}{}& BH, no condensate  &  inter \\
			\hline
			\multicolumn{1}{|c|}{\multirow{4}{*}{$c\neq 0$}} & \textbf{Type 3} & \textbf{Type 4}\\
			\multicolumn{1}{|c|}{}& un-joined, $\psi(r)$~$r$-dependent   & joined, $\psi(r)$~$r$-dependent  \\
			\multicolumn{1}{|c|}{}& Mink ($q=0$) intra &  intra+inter\\
			\multicolumn{1}{|c|}{}& BH ($q\ne0$) intra & only~when~$q\neq 0$\\
				\hline
		\end{tabular}}
		\end{center}	
	\caption{Types of possible solutions for the balanced charge $(q,-q)$ case, where (Mink,BH) stand for (Minkowski,black-hole) embeddings.}
	\label{typessolqnotzero}
\end{table}
Type 1 solutions  are maximally symmetric with $\psi=\frac{\pi}{2}$ and $z_{1,2}=\pm L/2$. They occur in the white region
of figure \ref{phase}.  They have no exciton condensates at all. 
Type 3 solutions occur in the red region.  They have $\psi(r)$ a nontrivial function, but $z_{1,2}=\pm L/2$.  The branes
do not join.
They are Minkowski embeddings when $q=0$ and black hole embeddings when $q\ne 0$. 
Type 3 has an intra-layer exciton condensate only.
There is a quantum phase transition between type 1 and type 3 solutions at $ q_c=0.0377$.
Both type 1 and type 3 solutions occur only for very small layer separations,  or order the UV 
cutoff.   
Type 2 solutions occupy the blue region.  They have  $\psi=\frac{\pi}{2}$, constant, $z_{1,2}(r)$ are 
nontrivial functions. The D7 and anti-D7 branes join 
at a radius, $r_0\neq 0$.  
The intra-layer condensate vanishes and there is a non-zero inter-layer condensate. 
In type 4 solutions, both $\psi(r)$ and $z_{1,2}(r)$ have nontrivial profiles. The D7 and anti-D7 branes join and
 $\psi(r)$ also varies with radius.  This phase has both and inter- and intra-layer condensate.
This solution exists only when $q$ is nonzero and, then, only for small values of $q$. 
For $r_0\simeq 0$ we have  $q<0.0377$, when $r_0$ grows, the allowed values of $q$ decrease. 

 Consider a double monolayer  with un-balanced charges, $Q>0$ on the D7  and $-\bar Q<0$ on the anti-D7 brane.
  The same argument as to why a single charged
  D7 brane must have a Minkowski embedding applies and, on the face of it, it is impossible for the branes to join before they
  reach the   horizon. There is, however, another possibility which arises when there are more than one species of
 fermions on each brane, that is, $N_F>1$. In that case, one or more of the fermion species can nest spontaneously,
  with the deficit of charge residing in the other species.  
  This would break internal symmetry.  For example, if $Q>\bar Q>0$, $k$ branes take up charge $\bar Q$ and the remaining 
  $N_F-k$ take up the remainder $Q-\bar Q$,   this would break $U(N_F)\times U(N_F)\to 
  U(N_F-k)\times U(k)\times U(N_F-k)\times U(k)$.  Then the branes with matched charges ($\bar Q$) would join, further
  breaking the symmetry $U(N_F-k)\times U(k)\times U(N_F-k)\times U(k)\to U(N_F-k)\times U(k)\times U(N_F-k)$.    
  Then, $N_F-k$ charged D7 branes and $N_F-k$ uncharged anti-D7 branes either break parity or
  some of the remaining $U(N_F-k)\times U(N_F-k)$ symmetry. 
  The uncharged branes must take up a Minkowski embedding.  
  We have computed the energies of some of these symmetry breaking states  for the case where $N_F=2$.   
  We find a range of charge densities where spontaneous nesting is 
  energetically preferred.  The implications of this idea for double monolayer physics is clear. 
  The fermion and hole 
  densities of individual monolayers would not necessarily have to be fine tuned in order to nest the Fermi surfaces.  
  It could happen spontaneously. 
    
  The intra-layer and inter-layer condensates   discussed here have not been seen in graphene to date
  (with a possible exception \cite{gordon}), presumably because the coupling is not strong enough. 
  Our results show that the inter-layer condensation is extremely sensitive to   Fermi surface nesting, even in 
  the strong coupling limit.  It would be interesting to better explore spontaneous nesting, since creating
  favourable conditions for it could be a way forward with graphene. 
 
 \noindent
  The work of G.W.S and N.K is supported by NSERC of Canada.

\end{document}